# Image Compression Using Proposed Enhanced Run Length Encoding Algorithm

A. H. Husseen, S. Sh. Mahmud, R. J. Mohammed

**Department of Computer Science ,College of Education Ibn Al- Haitham , University of Baghdad**

## Abstract

In this paper, we will present proposed enhance process of image compression by using RLE algorithm. This proposed yield to decrease the size of compressing image, but the original method used primarily for compressing a binary images [1].Which will yield increasing the size of an original image mostly when used for color images. The test of an enhanced algorithm is performed on sample consists of ten BMP 24-bit true color images, building an application by using visual basic 6.0 to show the size after and before compression process and computing the compression ratio for RLE and for the enhanced RLE algorithm.

## Introduction

The size of the compressed stream depends on the complexity of the image[2]. Image compression is minimizing the size in bytes of a graphics file without degrading the quality of the image to an unacceptable level. The reduction in file size allows more images to be stored in a given amount of disk or memory space. It also reduces the time required for images to be sent over the Internet or downloaded from Web pages[3]. Computer graphics applications, particularly those generating digital photographs and other complex color images, can generate very large file sizes. Issues of storage space, and the requirement to rapidly transmit image data across networks and over the Internet, have therefore led to the development of a range of image compression techniques, to reduce the physical size of files. Most compression techniques are independent of specific file formats – indeed, many formats support a number of different compression types.

**Graphics compression algorithms**

Graphics compression algorithms fall into two categories:

a- Lossy compression achieves its effect at the cost of a loss in image quality, by removing some image information.

b- Lossless compression techniques reduce size whilst preserving all of the original image information, and therefore without degrading the quality of the image. Although lossy techniques may be very useful for creating versions of images for day-to-day use or on the Internet, they should be avoided for archival master versions of image[4]

**Different formats of images**

Digital images can be stored in different formats. JPEG is a form developed by the Joint Photographic Expert Group. Images in this form compressed to a high degree. JPEG's purpose is to achieve high compression ratio with images containing large number of colors.

Graphics Interchange Format (GIF) is an 8-bit per pixel format. It supports animas well. GIF images are compressed using lossless LZW data compression technique. Portable Network Graphics (PNG) is another lossless data compression which is close to GIF but supports 24-bit color palette. It can be seen that JPEG can be used whenever the size is a criterion [5].

**JPEG:** Is a lossy compression scheme for color and gray-scale images. It works on full 24-bit color (True Color) which means it can store up to 16 million colors [6]. JPEG is favored because its compression produces a reasonable image with a small file size. JPEG's compression reduces image files to about 5% their normal size with loss due to destructive or "lossy" compression [7].

**TIFF:** The TIFF (Tagged Image File Format) format, it is a flexible format that normally saves 8 bits or 16 bits per color (red, green, blue) for 24-bit and 48-bit totals, using the LZW compression algorithm for lossless storage(lossless, but is less effective for 24 bit color images). Most graphics programs that use TIFF do not compression. Consequently, file sizes are quite big [8]. The TIFF file format is ideal for producing and storing high-quality images that are used to produce professional photographic prints, exhibit backgrounds, posters, magazines, etc[7].

**GIF:** Works best for images with only a few distinct colors, such as line drawings and simple cartoons. GIF is useful for cartoon images that have less than 256-(2^8) colors, grayscale images, and black and white text. The primary limitation of a GIF is that it only works on images with 8 bits per pixel or less, which means 256 or fewer colors. GIF compresses images using LZW compression [9]. GIF achieves compression in two ways. First, it reduces the number of colors of color-rich images, thereby reducing the number of bits needed per pixel, as just described. Second, it replaces commonly occurring patterns (especially large areas of uniform color) with a short abbreviation: instead of storing "white, white, white, white, white," it stores "5 white." Thus, GIF is "lossless" only for images with 256 colors or less. For a rich, true color image, GIF may "lose" 99.998% of the colors [10].

**BMP:** Windows BMP is the native image format in the Microsoft Windows operating systems. It supports images with 1, 4, 8, 16, 24, and 32 bits per pixel, although BMP files using 16 and 32 bits per pixel are rare. BMP also supports simple run-length compression for 4 and 8 bits per pixel. However, BMP compression is of use only with large blocks with identical colors, making it of very limited value. It is rare for Windows BMP to be in a compressed format. Multi-byte integers in the Windows BMP format are stored with the least significant bytes first. Data stored in the BMP format consists entirely of complete bytes so bit string ordering is not an issue. The BMP file structure is very simple [12]. It starts with a file header that contains the two bytes BM and the file size. This is followed by an image header with the width, height, and number of bitplanes[2]. Windows BMP files begin with a 54-byte header [13]:

| Offset | Size | Description |
|---|---|---|
| 0 | 2 | signature, must be 4D42 hex |
| 2 | 4 | size of BMP file in bytes (unreliable) |
| 6 | 2 | reserved, must be zero |
| 8 | 2 | reserved, must be zero |
| 10 | 4 | offset to start of image data in bytes |
| 14 | 4 | size of BITMAPINFOHEADER structure, must be 40 |
| 18 | 4 | image width in pixels |
| 22 | 4 | image height in pixels |
| 26 | 2 | number of planes in the image, must be 1 |
| 28 | 2 | number of bits per pixel (1, 4, 8, or 24) |
| 30 | 4 | compression type (0=none, 1=RLE-8, 2=RLE-4) |
| 34 | 4 | size of image data in bytes (including padding) |
| 38 | 4 | horizontal resolution in pixels per meter (unreliable) |
| 42 | 4 | vertical resolution in pixels per meter (unreliable) |
| 46 | 4 | number of colors in image, or zero |
| 50 | 4 | number of important colors, or zero |

**PNG:** Provide a non-proprietary alternative to the LZW compression employed by GIF and other file formats. PNG compression uses the deflate compression method. It is lossless algorithm and is effective with color depths from 1-bit (monochrome) to 48-bit (True color). It allows you to make a trade-off between file size and image quality when the image is compressed. Typically, an image in a PNG file can be 10 to 30% more compressed than in a GIF format[4] . PNG uses ZIP compression which is lossless, and slightly more effective than LZW (slightly smaller files)[11].

**Run length encoding (RLE)**

Run length encoding (RLE) is perhaps the simplest compression technique in common use. RLE algorithms are lossless, and work by searching for runs of bits, bytes, or pixels of the same value, and encoding the length and value of the run. As such, RLE achieves best results with images containing large areas of contiguous color, and especially monochrome images. Complex color images, such as photographs, do not compress well – in some cases, RLE can actually increase the file size. There is a number of RLE variants in common use, which are encountered in the TIFF, PCX and BMP graphics formats [4]. Run-length encoding represents a string by replacing each subsequence of consecutive identical characters with(char; length). The string 11112222333111 would have representation (1; 4)(2; 4)(3; 3)(1; 3). Then compress each (char; length) as a unit using, say, Human coding. Clearly, this technique works best when the characters repeat often. One such situation is in fax transmission, which contains alternating long sequences of 1's and 0's. The distribution of code words is taken over many documents to compute the optimal Human code.[14]

**Proposed enhanced run length encoding algorithm**

Each color images consists of the basic three colors (R,G,B), the RLE algorithm represented the image consists on N pixels as follows :

| Red R | Green G | Blue B | Number of pixels C |
|---|---|---|---|
| $r_1$ | $g_1$ | $b_1$ | $c_1$ |
| $r_2$ | $g_2$ | $b_2$ | $c_2$ |
| …… | …… | …… | …… |
| …… | …… | …… | …… |
| $r_n$ | $g_n$ | $b_n$ | $c_n$ |

But in the new algorithm we compute the differences between the adjacent pixels for each color, if the difference between $r_1$ and $r_2$ less than or equal to a threshold value (th<=10) and if the difference between $g_1$ and $g_2$ less than or equal to a threshold value (th<=10) and if the difference between $b_1$ and $b_2$ less than or equal to a threshold value (th<=10) we add 1 to $C_1$, and if the difference is greater than 10 we do this process between next adjacent pixels until we reach the last pixel in the image, and the following example explains the RLE algorithm and the enhanced RLE algorithm **:**

Let we have the image of 4x4 pixels

| 100 | 101 | 102 | 100 |
|---|---|---|---|
| 200 | 200 | 205 | 209 |
| 300 | 300 | 305 | 301 |
| 210 | 205 | 300 | 300 |

The number of records required to save this image are 16. When compress this image by using RLE algorithm we will generate the following values :

100(1) , 101(1), 102(1), 100(1), 200(2), 205(1), 209(1), 300(2), 305(1), 301(1), 210(1), 205(1), 300(2).

The number of records required to save this image are 13. But when compress this image by using an enhanced RLE algorithm we will generate the following values 100(4) , 200(4), 300(4), 210(2), 300(2).

The number of records required to save this image are 5 .

**The proposed enhanced RLE algorithm**

1- Read the BMP image.

2- Get the height N and the width M for the image

3- Create an array , let it Main(N,M) ,each element of this array consists of three fields for R,G,B colors.

4- Convert the image to the main array ; Main(N,M).

5- Let X=Main(0,0) ; Main(0,0) is the first element in an array.

    Let TH=10, TH : the threshold .

6- For I= 0 to N-1

7- For J=0 to M-1

8-     If X-Main(I,J) <= TH then

        Let C=C+1

    Else

        Let X=Main(I,J) and C=0

9- End.

**Experimental results**

we use ten images to test the above algorithm and the threshold value equal or less than (10), then we get the following results :

| Image Name (BMP) | Original Size (KB) | New Size (KB) Using RLE | Compression Ratio Original Size/ Size Using RLE | New Size (KB) Using proposed enhanced RLE | Compression Ratio Original Size/ Size Using proposed enhanced |
|---|---|---|---|---|---|
| Birds | 475 | 586 | 0.8:1 | 133 | 3.5:1 |
| Heart | 919 | 1137 | 0.86:1 | 471 | 1.95:1 |
| Light | 226 | 298 | 0.003:1 | 211 | 1.07:1 |
| Madina | 80 | 89 | 0.9:1 | 38 | 2.1:1 |
| Mansour | 88 | 116 | 0.98:1 | 23 | 3.8:1 |
| Microsoft | 523 | 526 | 0.99:1 | 145 | 3.6:1 |
| Ramadan | 533 | 669 | 0.79:1 | 240 | 2.2:1 |
| Rawda | 998 | 775 | 1.28:1 | 343 | 2.9:1 |
| Rose | 127 | 160 | 0.79:1 | 62 | 2.08:1 |
| Winnie | 170 | 64 | 2.65:1 | 31 | 5.48:1 |

The above table shows the size of true color BMP images and the size after the compression process using RLE and proposed enhanced RLE algorithms.

The compression ratio is changing from image to another, but this related to the variety between the values of adjacent pixels.

Decreasing variety between the values of adjacent pixels will yield to increase compression ratio, and vice versa. As shown in the figures (4, 5, 6, 7, 8, 9, 10, 11, 12, 13).

The table below appears the size of all sample images when converted to JPEG, GIF, PNG, TIFF types using Microsoft Paint:

| Image Name (BMP) | Original Size (KB) | New size (KB) | | | | |
|---|---|---|---|---|---|---|
| | | proposed enhanced RLE | JPEG format | GIF format | TIFF format | PNG format |
| Birds | 475 | 133 | 20 | 43 | 361 | 247 |
| Heart | 919 | 471 | 54 | 123 | 958 | 686 |
| Light | 226 | 211 | 25 | 31 | 239 | 183 |
| Madina | 80 | 38 | 6 | 6 | 23 | 19 |
| Mansour | 88 | 23 | 5 | 10 | 84 | 68 |
| Microsoft | 523 | 145 | 18 | 52 | 434 | 285 |
| Ramadan | 533 | 240 | 37 | 54 | 485 | 350 |
| Rawda | 998 | 343 | 38 | 90 | 564 | 349 |
| Rose | 127 | 62 | 8 | 17 | 131 | 99 |
| Winnie | 170 | 31 | 8 | 9 | 52 | 40 |

## Conclusions

1- Run length encoding algorithm is a method of compressing images depend on the number of adjacent pixels value in the image.
2- RLE algorithm is failing mostly when using it for compressing color images, because we need new field to store the number of repeated adjacent pixels.
3- Run length encoding algorithm is an efficient compression method with images have less various between values of adjacent pixels, but fail with images have high difference between adjacent pixels value.
4- When increasing the values of threshold in the proposed enhanced RLE algorithm this will yield increasing the compression ratio, and vice versa.
5. Controlling the compression ratio depends on the value of the threshold, which depends on the type of domain that image used it .

## References

1. Umbaugh,S.E. (1998), Computer Vision and Image Processing: A Practical Approach Using CVIPtools, First Edition, Prentice Hall, New York,253.
2. Salomon, D. (2007), Data Compression", Fourth Edition, Springer-Verlag,London,28-36.
3. http://searchsoa.techtarget.com/definition/imagecompression
4. Adrian, B. (9 July 2003), Image compression, Digital Preservation Guidance note 5, DPGN-05, 1-9,www.nationalarchives.gov.uk/documents/image_compression.pdf
5. Srinivasan,V., Image Compression Using Embedded Zero-tree Wavelet Coding (EZW),2, www.umiacs.umd.edu
6. Blelloch,E. (25 September 2010) , Introduction to Data Compression, Computer Science Department ,Carnegie Mellon University,www.cs.cmu.edu/afs/cs/project/pscico-guyb/realworld/www/compression.pdf
7. Carey,L. (August 2008), Understanding Digital Image, Conserve O Gram, Sponsored by the Park Museum Management Program, 22/2, 1-2, www.nps.gov/history/museum/publications/conserveogram/22-02.pdf
8. http://en.wikipedia.org/wiki/Image_file_formats
9. www.contentdm.com/USC/tutorials/image-file-types.pdf
10. www.wfu.edu/~matthews/misc/graphics/formats/formats.html
11. www.scantips.com/basics09.html
12. Miano, J. (1999), Compressed image file formats: JPEG, PNG, GIF, XBM, BMP, ACM Press/Addison-Wesley, New York.
13. www.fastgraph.com/help/bmp_os2_header_format.html
14. Jessica, F. (12 February 2002), "Algorithms for Massive Data Sets Context-based Compression", CS 493,1:1-3,www.docstoc.com/docs/54164044/Run-length-encoding

# عملية ضغط الصور باستعمال خوارزمية الـ RLE المطورة

**علي هادي حسين، شيماء شاكر محمود، رولا جاسم محمد**

**قسم علوم الحاسبات ،كلية التربية ـ ابن الهيثم ، جامعة بغداد**

## الخلاصة

في هذه الدراسة سنعرض عملية تحسين على خوارزمية RLE المستعملة في عملية ضغط الصور، خوارزمية RLE تستعمل بكفاية لضغط الصور الثنائية Binary Images [1]، عند استعمال خوارزمية RLE مباشرة لضغط الصور الملونة True color فإن ذلك يؤدي في الأغلب إلى زيادة حجم الصورة المضغوطة عن الصورة الأصلية. اختبرت الخوارزمية الجديدة على عينة تتألف من عشر صور ملونة نوع BMP 24-bit True Color وبناء تطبيق بلغة Visual Basic 6.0 وبيان الحجم الأصلي للصور والحجم الجديد بعد عملية الضغط وحساب نسبة الضغط Compression Ratio باستعمال خوارزمية RLE الاعتيادية والمحسنة.

**الكلمات المفتاحية** : عملية الضغط ، RLE، Run length encoding، GIF، TIFF، JPEG PNG، BMP ، هيكل BMP، ملف BMP ، نسبة الضغط ، فقدان ، بدون فقدان ، True color.

**Appendix for Application Form and all Images used as a sample**

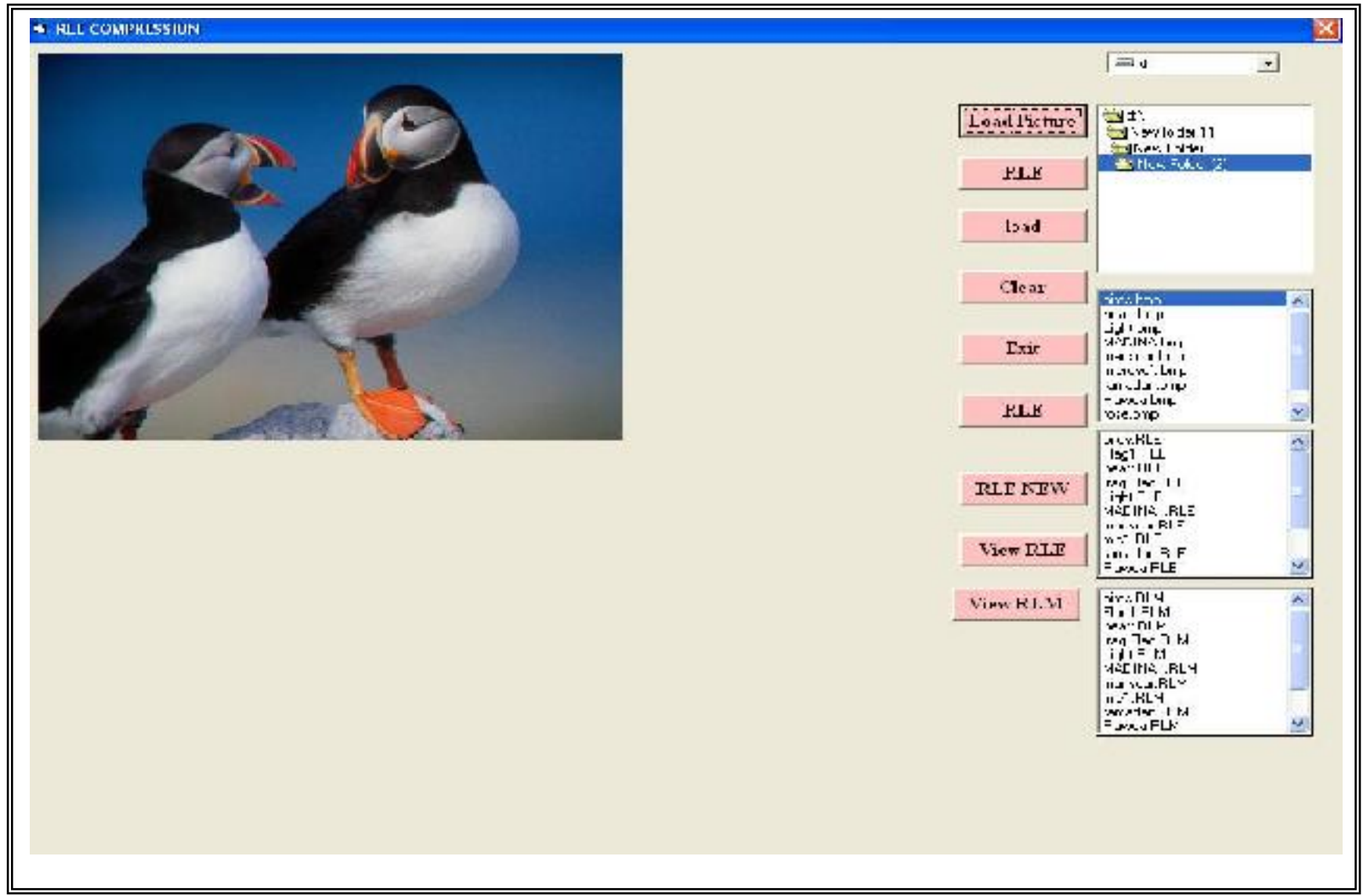

Fig. (1): Application form built using visual basic 6.0

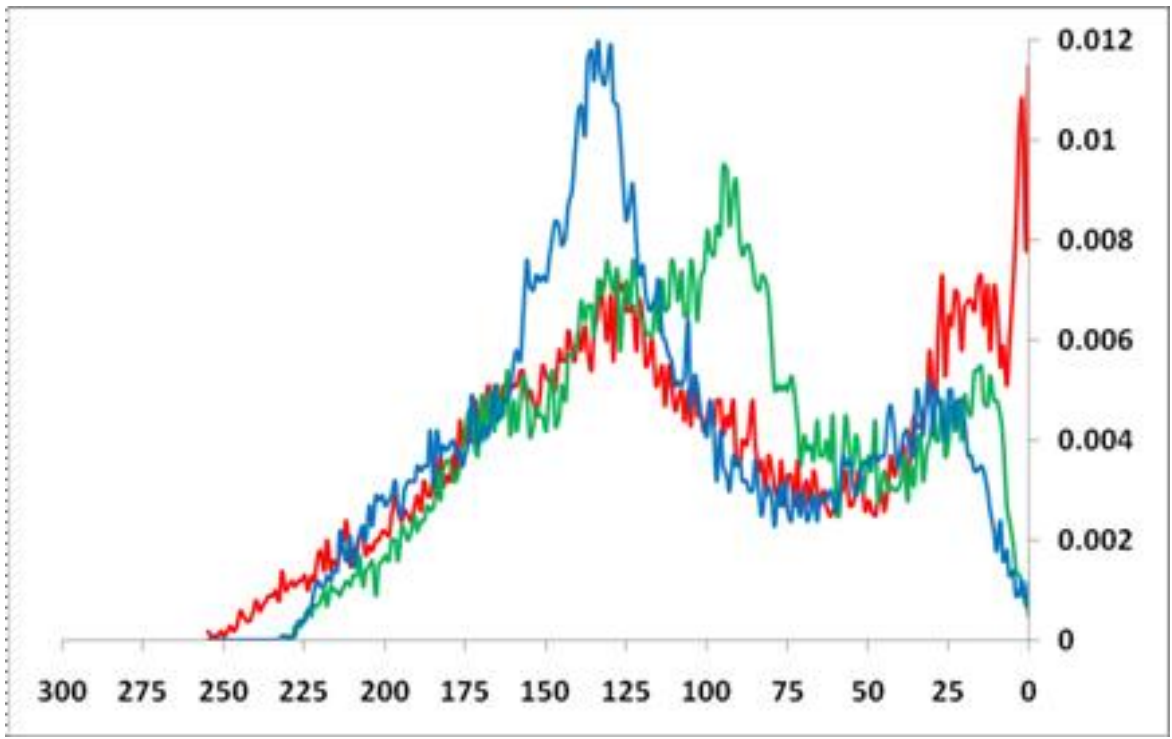

**Fig. (2): Histogram for the image Birds.BMP using RLE algorithm**

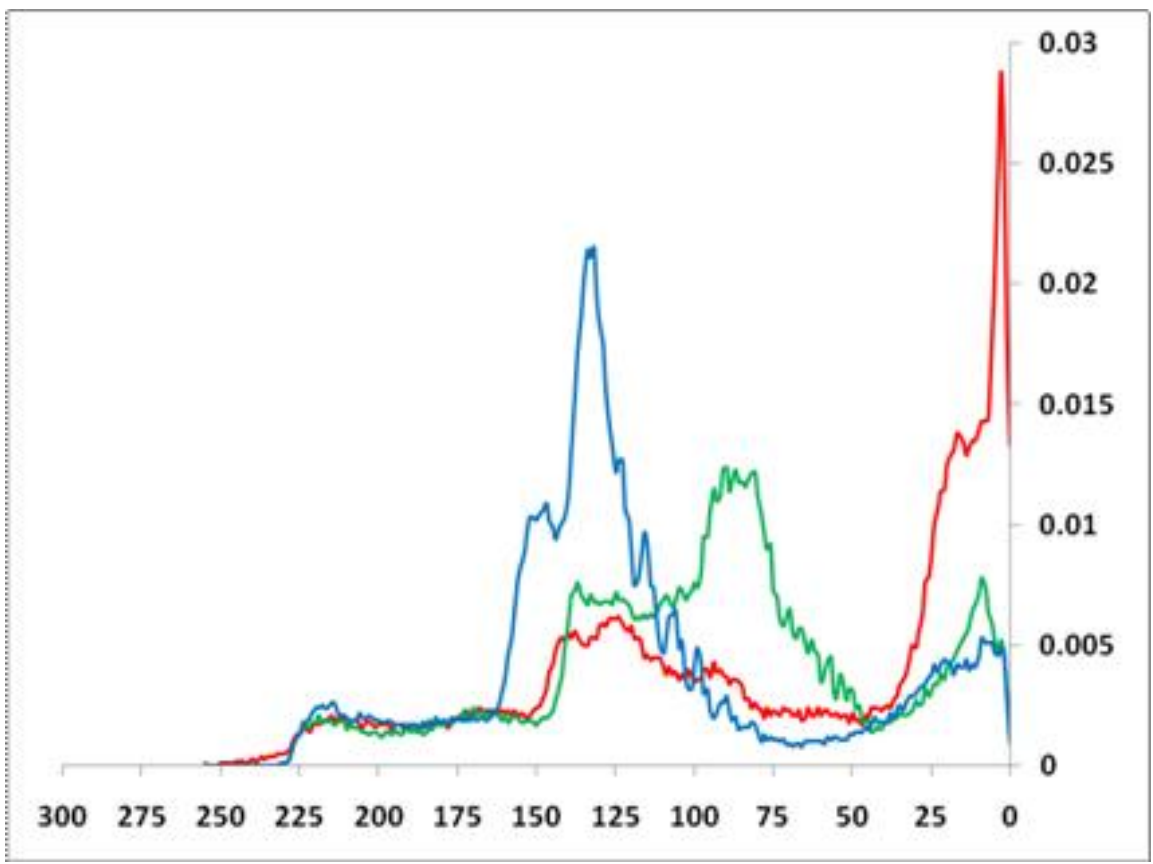

**Fig. (3): Histogram for the image Birds.BMP using proposed enhanced RLE algorithm**

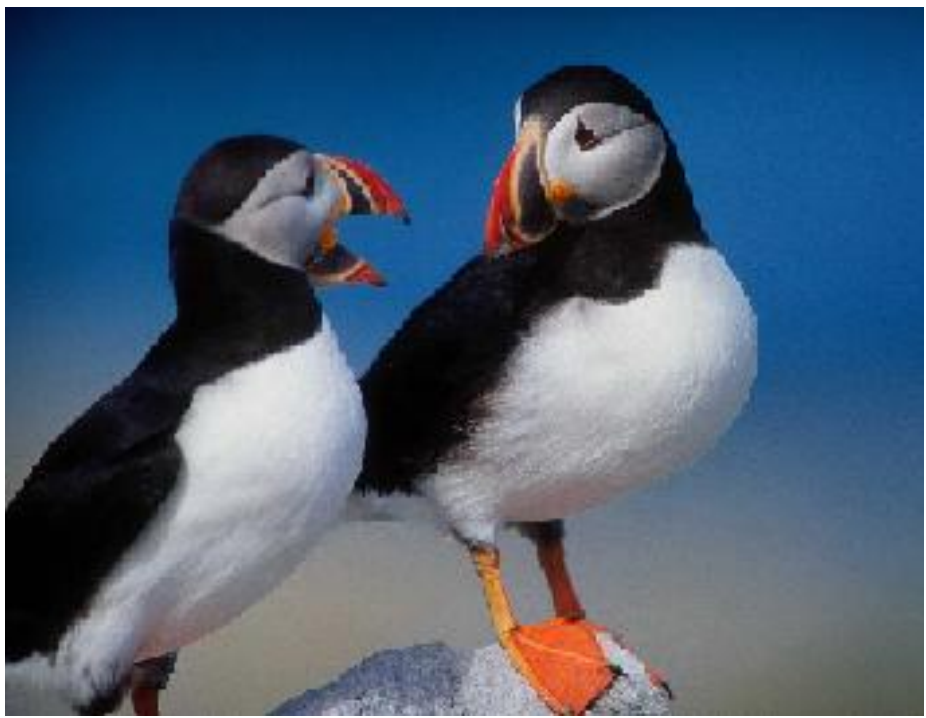

Fig. (4): Birds.BMP 456x355 pixels

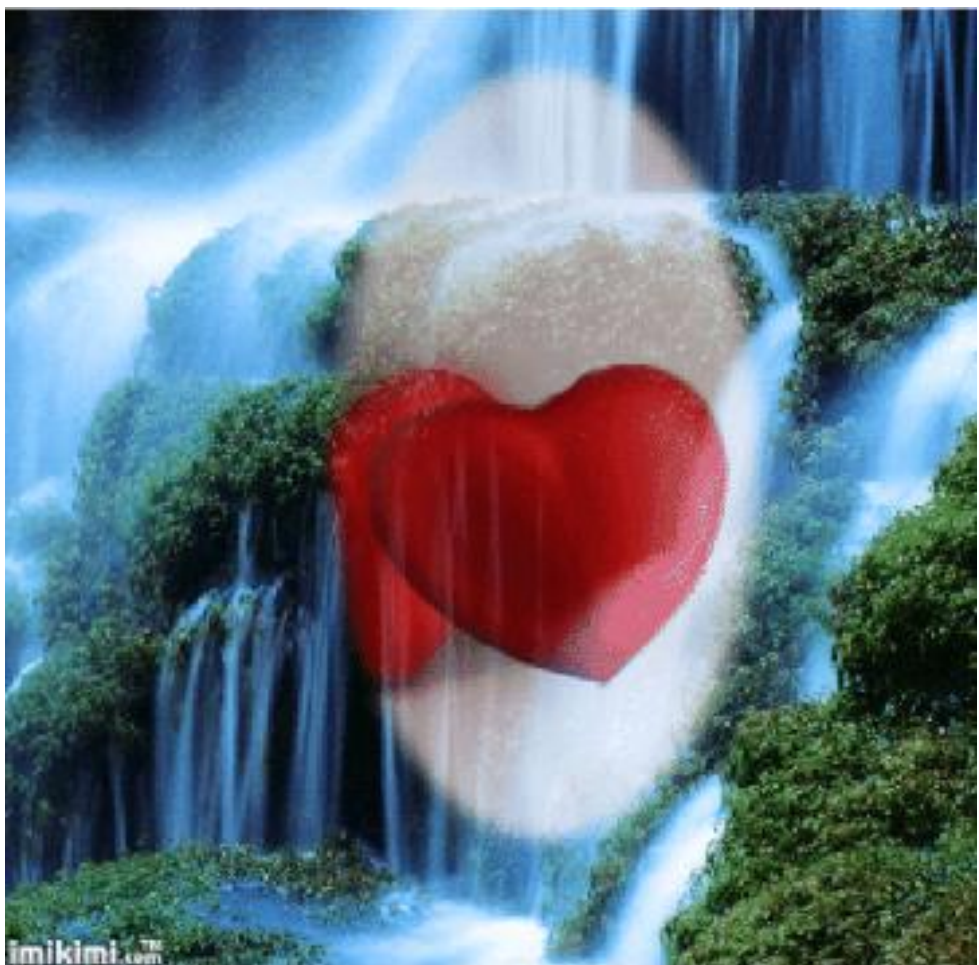

Fig. (5): Heart.BMP 560x560 pixels

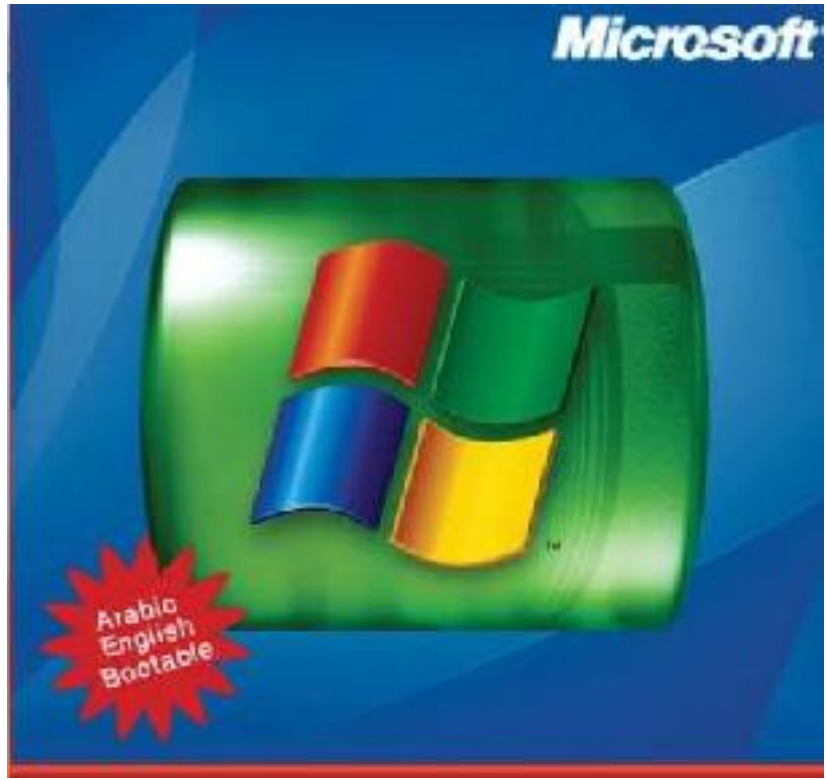

**Fig .(6): Microsoft.BMP 432x413 pixels**

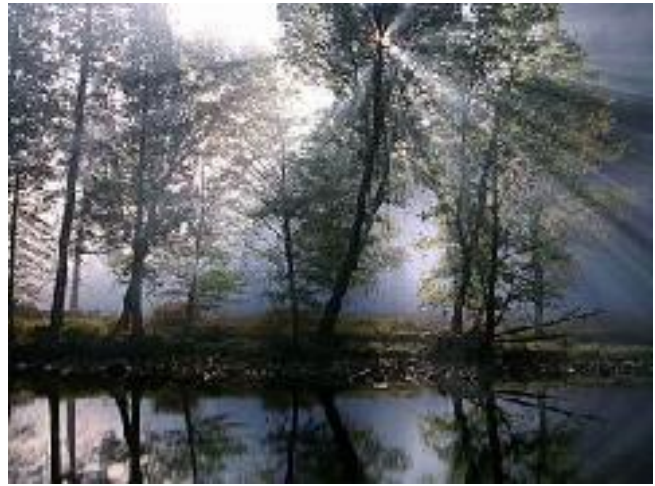

**Fig .(7): Light.BMP 320x240 pixels**

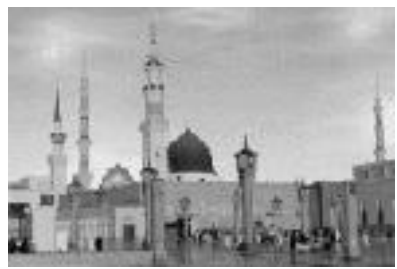

**Fig. (8): Madina.BMP 200x135 pixels**

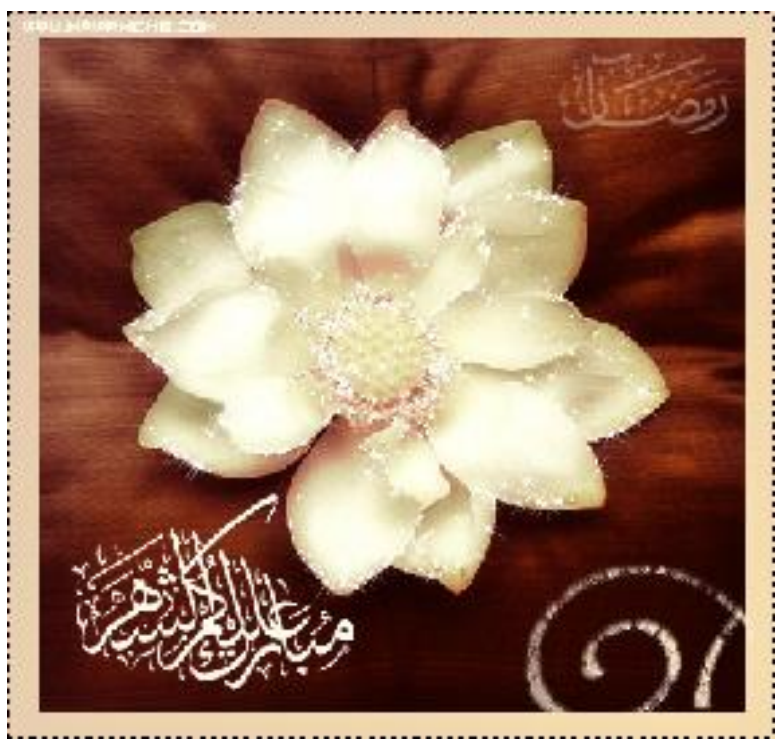

**Fig. (9): Ramadan.BMP 436x417 pixels**

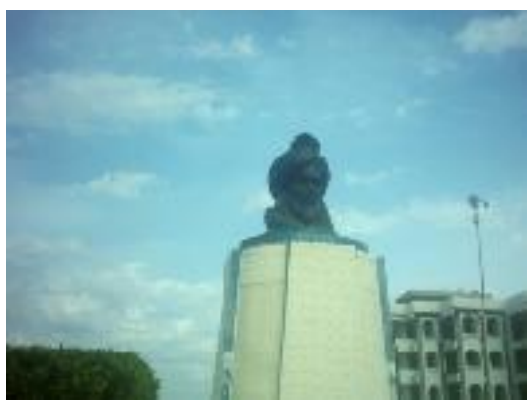

**Fig .(10): Mnsour.BMP 200x150 pixels**

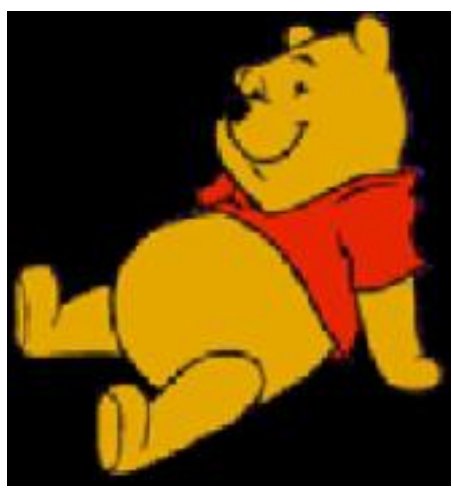

**Fig. (11): Winnie.BMP 232x250 pixels**

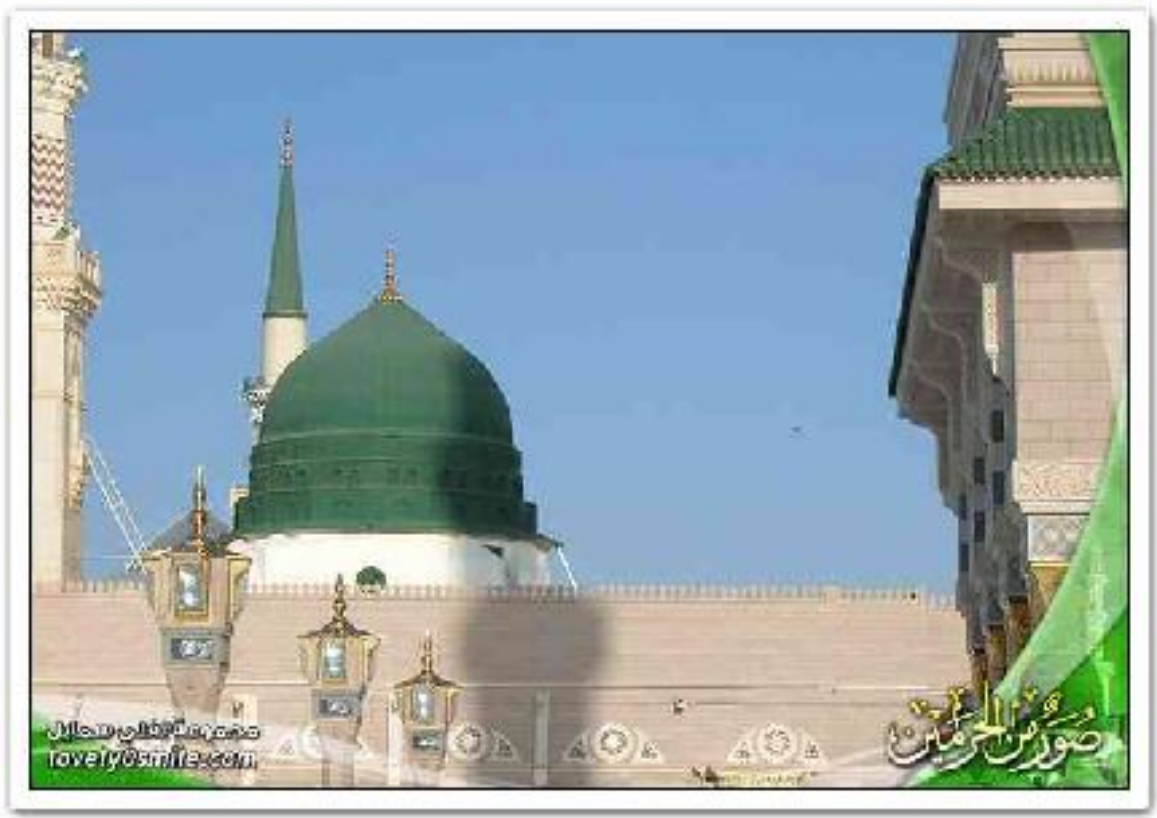

**Fig. (12): Rawda.BMP 692x492 pixels**

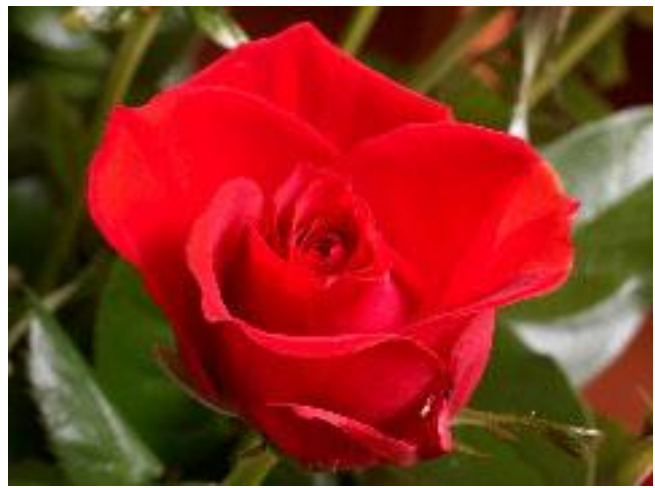

**Fig .(13): Rose.BMP 240x180 pixels**